\documentclass{ifacconf}

\usepackage{graphicx}      % include this line if your document contains figures
\usepackage{natbib}        % required for bibliography
\usepackage{amsmath}
\usepackage{xcolor}
\usepackage{pdflscape}
\usepackage{amssymb}
\usepackage{subcaption}

\newtheorem{theorem}{Theorem}[section]
\newtheorem{lemma}[theorem]{Lemma}
\newtheorem{remark}[theorem]{Remark}
\newtheorem{assumption}[theorem]{Assumption}

\allowdisplaybreaks

\begin{document}
\begin{frontmatter}

\title{Distributed Observer-based Leak Detection in Pipe Flow with Nonlinear Friction\\ (Extended Abstract)}


%\thanks[footnoteinfo]{Sponsor and financial support acknowledgment
%goes here. Paper titles should be written in uppercase and lowercase
%letters, not all uppercase.}

\author[First]{Nils Christian A. Wilhelmsen} 
\author[First]{Ole Morten Aamo} 
%\author[Third]{Third C. Author}

\address[First]{Department of Engineering Cybernetics, Norwegian University of Science and Technology, NO-7491 Trondheim, Norway (e-mail: nils.c.wilhelmsen@ntnu.no, aamo@ntnu.no).}
%\address[First]{Colorado State University, 
%   Fort Collins, CO 80523 USA (e-mail: author@lamar. colostate.edu)}
%\address[Third]{Electrical Engineering Department, 
 %  Seoul National University, Seoul, Korea, (e-mail: author@snu.ac.kr)}

\begin{abstract}                % Abstract of not more than 250 words.
The problem of leak detection in a pipeline with nonlinear friction is considered. A distributed observer-based method is proposed which applies a linearised, distributed adaptive observer design to the nonlinear model. The methodology is tested in simulations for two different operating points.
\end{abstract}


\end{frontmatter}
%===============================================================================

\section{Introduction}

Leakage in pipelines is a ubiquitous problem in water distribution systems~(\cite{lambert2002international}), which motivates the development of technology to monitor and automatically detect leaks as they occur. Software-based methods relying on sophisticated processing of sensor signals to detect and quantify leaks have been studied over the past few decades~(\cite{billmann1987leak}, \cite{verde2001multi}).

In~\cite{aamo2015leak} an adaptive observer-based leak detection scheme for a distributed pipeline model is designed using the infinite-dimensional backstepping technique. This design has later been extended to more general cases, such as branched pipe flows in~\cite{anfinsen2022leak} and pipe flow networks containing loops in~\cite{wilhelmsen2022leak}. These contributions have, however, only considered linear models of the pipe flow in their design and implementation. In this extended abstract we explore the possibility of applying the leak detection scheme from~\cite{aamo2015leak} to a pipeline model with nonlinear friction, something which could be of great practical value.

\section{Mathematical model}

\subsection{Nonlinear model}
To model the transient behaviour of pressure $p$ and volumetric flow $q$ at each position $z \in [0,l]$ and time $t > 0$ in a pipe of length $l$ we consider the $1D$ semilinear pipeline flow model~(see e.g.~\cite{aamo2006observer})
\begin{subequations} 
\begin{align}
	p_t(z,t) + \frac{\beta}{A}q_z(z,t)  &= - \frac{\beta}{A}d(z)\chi\\
	q_t(z,t) + \frac{A}{\rho}p_z(z,t) &=  - Ag \sin \phi(z)  - \frac{\eta}{A}d(z)\chi  \notag \\ & \quad -\frac{f}{2}\frac{|q(z,t)|q(z,t)}{DA} 
\end{align}
\label{eq:semilinear_pipe_model}
\end{subequations}
%\begin{subequations} \label{eq:nonlinear_pipe_model}
%\begin{align}
%	p_t(z,t) + \frac{1}{A}q(z,t)p_z(z,t)  &= -\frac{\beta}{A}q_z(z,t) - \frac{\beta}{A}d(z)\chi\label{eq:mass_balance}\\
%	q_t(z,t) + \frac{1}{A}q(z,t)q_z(z,t)  &=  -\frac{A}{\rho}p_z(z,t) - Ag \sin \phi(z)  \notag \\ & \quad -\frac{f}{2}\frac{|q(z,t)|q(z,t)}{DA} \notag \\ & \quad - \frac{\eta}{A}d(z)\chi\label{eq:momentum_balance}
%\end{align}
%\label{eq:quasilinear_pipeline}
%\end{subequations}
with boundary conditions
\begin{subequations}
\begin{align}
q(0,t) &= q_0(t)\\
p(l,t) &= p_l(t).
\end{align}
\label{eq:pipeline_boundary}
\end{subequations}

Possible leaks are characterised by the total volumetric flow $\chi$ leaking from the pipeline together with their normalised\footnote{It is imposed that $\int_0^l d(z)dz = 1$.} spatial distribution $d(z) = \delta''(z)$ throughout the pipeline, where
\begin{equation}
\delta(z) := l-z-\int_z^l\int_0^{\eta}d(\gamma)d\gamma d\eta.
\end{equation}
 
 Also, $A$ is the cross-sectional area of the pipeline, $D$ is the diameter, $\beta$ is the bulk modulus, $\rho$ is the density~(assumed constant), $g$ is the acceleration of gravity, $\phi$ is the pipeline inclination angle and $f$ is a friction coefficient satisfying
\begin{equation}
	\frac{1}{\sqrt{f}} = -1.8log_{10}\left( \left(\frac{E}{3.7D}\right)^{1.11} + \frac{6.9}{Re} \right)
\end{equation} 
in terms of the pipe surface roughness $E$ and Reynolds number
\begin{equation}
Re := \frac{\rho q D}{A\mu},
\label{eq:Reynolds}
\end{equation}
with $\mu$ being the dynamic viscosity. Lastly the parameter $\eta := \gamma_dq_{in}$ is defined in terms of the mean flow $q_{in}$ and coefficient $\gamma_d < 1$ characterising the leak type~(\cite{bajura1971model}), with $\gamma_d \approx 0.8$ for point leaks.

\subsection{Problem statement}
Our objective is to monitor flow in pipelines with dynamics described by~\eqref{eq:semilinear_pipe_model}--\eqref{eq:pipeline_boundary} through providing estimates of the pressure $p$ and volumetric flow rate $q$ in addition to detection with size and location estimates of possible leaks that occur in the pipeline. We will assume that flow and pressure measurements from the boundaries at $z=0$ and $z=l$ are available, only.

In~\cite{aamo2015leak} a distributed observer-based leak detection method using only boundary pressure and flow measurements is proposed for pipelines described by~\eqref{eq:semilinear_pipe_model}--\eqref{eq:pipeline_boundary}, but assuming the dynamics~\eqref{eq:semilinear_pipe_model} are linearised. Rigorous proofs are given for convergence of the leak size estimate $\hat \chi(t) \rightarrow \chi$ as well as leak distribution estimate $\hat \delta(t) \rightarrow \delta(0)$. %, where
%\begin{equation}
%\delta(z) := l-z-\int_z^l\int_0^{\eta}d(\gamma)d\gamma d\eta.
%\end{equation}
The assumption that the friction term can be linearised is only valid for laminar flow~(\cite{moody1944friction}), which for pipes is characterised by ``low'' Reynolds number $Re < 2300$. In practice, however, applications such as water distribution systems contain flow with higher Reynolds numbers in the turbulent regime. The nonlinearity of the friction term can in these cases thus not be neglected and must be taken into account in model-based leak detection methods. It is therefore of interest to consider how well the linearised observer from~\cite{aamo2015leak} works when applied to a dynamic model such as~\eqref{eq:semilinear_pipe_model} with nonlinear terms. 

The linear observer from~\cite{aamo2015leak} uses observer gains computed based on a given operating point $(q_{in}, p_{out})$ the input signals $q_0, p_l$ in the boundary condition~\eqref{eq:pipeline_boundary} are assumed to fluctuate around. For water distribution systems, these operating points are in reality slowly~(relative to the time scale of the transient dynamics~\eqref{eq:semilinear_pipe_model}) changing based on both seasonal variations but also differences in water demand between day and night~(\cite{maidment1986daily}). Thus for implementation the observer gains need to be updated at regular intervals to correspond to predicted demand levels.
\iffalse
As considered in~\cite{aamo2006observer}, the eigenvalues of~\eqref{eq:quasilinear_pipeline} are given by
\begin{align}
\lambda_1(q) = \frac{q}{A} - c, &&  \lambda_2(q) = \frac{q}{A} + c
\end{align}
with $c = \sqrt{\frac{\beta}{\rho}}$ the speed of sound. In practice, for water distribution systems, the water typically flows at speeds $\frac{q}{A} << c$, and we can thus approximate $\lambda_1 \approx  - c$, $\lambda_2 \approx  c$. Hence the terms $\frac{1}{A}qp_z$, $\frac{1}{A}qq_z$ in~\eqref{eq:quasilinear_pipeline} are neglected. 

Using that $d(z) = \delta''(z)$, our objective is to modify and test the leak detection method from~\cite{aamo2015leak} for the dynamics
\begin{subequations} 
\begin{align}
	p_t(z,t) + \frac{\beta}{A}q_z(z,t)  &= - \frac{\beta}{A}\delta''(z)\chi\\
	q_t(z,t) + \frac{A}{\rho}p_z(z,t) &=  - Ag \sin \phi(z)  - \frac{\eta}{A}\delta''(z)\chi  \notag \\ & \quad -\frac{f}{2}\frac{|q(z,t)|q(z,t)}{DA} 
\end{align}
\label{eq:semilinear_pipe_model}
\end{subequations}

\noindent with boundary conditions~\eqref{eq:pipeline_boundary}.\fi

 Our objective is hence to modify and test the leak detection method from~\cite{aamo2015leak} for the dynamics~\eqref{eq:semilinear_pipe_model}--\eqref{eq:pipeline_boundary}. The boundary signals $q_0$ and $p_l$ in~\eqref{eq:pipeline_boundary} in addition to $p_0 = p(0,\cdot)$ and $q_l = q(l,\cdot)$ are assumed measured. 

\subsection{Riemann coordinates}

To apply the observer from~\cite{aamo2015leak} to~\eqref{eq:semilinear_pipe_model}--\eqref{eq:pipeline_boundary}, we write it first in Riemann coordinates.
\begin{lemma}
The change of coordinates
\begin{subequations}
\begin{align}
u(x,t)  &:= \frac{1}{2}\Bigg(q(xl,t) + \delta'(xl)\chi + \frac{A}{\sqrt{\beta \rho}}\Bigg( p(xl,t) \notag \\ & \quad + \rho g \int_0^{xl}\sin \theta(\gamma)d\gamma + \frac{\rho}{A^2}\eta\delta'(xl)\chi  \Bigg) \Bigg)\\
v(x,t) &:= \frac{1}{2}\Bigg(q(xl,t) + \delta'(xl)\chi - \frac{A}{\sqrt{\beta \rho}}\Bigg( p(xl,t)\notag \\ & \quad + \rho g \int_0^{xl}\sin \theta(\gamma)d\gamma  + \frac{\rho}{A^2}\eta\delta'(xl)\chi  \Bigg) \Bigg)
\end{align}
\end{subequations}
maps~\eqref{eq:semilinear_pipe_model}--\eqref{eq:pipeline_boundary} into
\begin{subequations}
\begin{align}
u_t(x,t) &= -\frac{1}{l}\sqrt{\frac{\beta}{\rho}}u_x(x,t)\notag \\ & \quad - \frac{f}{4DA}|u(x,t)+v(x,t) - \delta'(xl)\chi|\notag \\ & \quad \times(u(x,t)+v(x,t) - \delta'(xl)\chi)\\
v_t(x,t) &= \frac{1}{l}\sqrt{\frac{\beta}{\rho}}v_x(x,t)\notag \\ & \quad - \frac{f}{4DA}|u(x,t)+v(x,t)- \delta'(xl)\chi|\notag \\ & \quad \times(u(x,t)+v(x,t)- \delta'(xl)\chi)\\
u(0,t) &= -v(0,t) + q_0(t) - \chi\\
v(1,t) &= \frac{1}{2}( q_l(t) - \frac{A}{\sqrt{\beta\rho}}(p_l(t) + \rho gh) ).
\end{align}
\label{eq:semilinear_plant}
\end{subequations}
\end{lemma}

\section{Observer}
\subsection{Linearisation}
Decomposing $(u,v)$ as perturbations $(\Delta \underline u, \Delta \underline v)$ around mean values $(\bar u, \bar v)$ so that
\begin{subequations}
\begin{align}
u(x,t) &\approx \bar u + \Delta \underline u(x,t)\\
v(x,t) &\approx \bar v + \Delta \underline v(x,t),
\end{align}
\label{eq:linearisation_decomposition}
\end{subequations}\\
\noindent and using that $q_{in} = \bar u + \bar v$, we have a linearisation of~\eqref{eq:semilinear_plant} given by
\begin{subequations}
\begin{align}
\Delta \underline u_t(x,t) &= -\frac{1}{l}\sqrt{\frac{\beta}{\rho}} \Delta \underline u_x(x,t)\notag \\ & \quad -\frac{f}{2DA}|q_{in}|(\Delta \underline u + \Delta \underline v - \delta'(xl)\chi)\\
\Delta \underline v_t(x,t) &= \frac{1}{l}\sqrt{\frac{\beta}{\rho}} \Delta \underline v_x(x,t)\notag \\ & \quad -\frac{f}{2DA}|q_{in}|(\Delta \underline u + \Delta \underline v - \delta'(xl)\chi)\\
\Delta \underline u(0,t) &= -\Delta \underline v(0,t) + q_0(t)- q_{in} - \chi\\
\Delta \underline v(1,t) &= \frac{1}{2}\left( q_l(t) - q_{in} - \frac{A}{\sqrt{\beta\rho}}(p_l(t) + \rho gh)\right).
\end{align}
\label{eq:linearised_plant}
\end{subequations}
We perform a final coordinate transformation
\begin{subequations}
\begin{align}
\Delta  \underline u &= \Delta u \exp(-\frac{lF}{2\sqrt{\beta\rho}}x) + \frac{A}{2\sqrt{\beta\rho}}(p_{out}  + \frac{F}{A}\delta(xl)\chi)\\
\Delta  \underline v &= \Delta v \exp(\frac{lF}{2\sqrt{\beta\rho}}x) - \frac{A}{2\sqrt{\beta\rho}}(p_{out} + \frac{F}{A}\delta(xl)\chi)
\end{align}
\label{eq:final_coordinate_transform}
\end{subequations}\\
with
\begin{equation}
F := \frac{f\rho}{DA}|q_{in}|
\end{equation}
to bring~\eqref{eq:linearised_plant} into the form
\begin{subequations}
\begin{align}
\Delta  u_t(x,t) &= -\epsilon \Delta u_x(x,t) + c_1(x)\Delta v(x,t)\\
\Delta  v_t(x,t) &= \epsilon \Delta  v_x(x,t) + c_2(x)\Delta u(x,t)\\
\Delta u(0,t) &= -\Delta  v(0,t) + q_0(t)- q_{in} - \chi\\
\Delta v(1,t) &= U(t)
\end{align}
\label{eq:linear_plant}
\end{subequations}
where we have used
\begin{align}
\epsilon &= \frac{1}{l}\sqrt{\frac{\beta}{\rho}}\label{eq:epsilon}\\
c_1(x) &= -\frac{F}{2\rho}\exp(\frac{lF}{\sqrt{\beta\rho}}x)\\
c_2(x) &= -\frac{F}{2\rho}\exp(-\frac{lF}{\sqrt{\beta\rho}}x)\\
U(t) &= \frac{1}{2}\exp(-\frac{lF}{2\sqrt{\beta\rho}}x)(q_l(t)-q_{in} - \frac{A}{\sqrt{\beta\rho}}(p_l(t)\notag \\ & \quad - p_{out} + \rho gh + \frac{Fl}{A}q_{in})).
\end{align}

\subsection{Linear observer}
The adaptive observer from~\cite{aamo2015leak} for~\eqref{eq:linear_plant} is given by
\begin{subequations}
\begin{align}
 \Delta \hat u_t(x,t)  &= -\epsilon \Delta \hat  u_x(x,t) + c_1(x) \Delta \hat v(x,t)\notag \\ & \quad + p_1(x)(\Delta y(t) - \Delta\hat u(1,t))\\
\Delta \hat v_t(x,t) &= \epsilon \Delta\hat v_x(x,t) + c_2(x)\Delta\hat u(x,t)\notag \\ & \quad  + p_2(x)(\Delta y(t) - \Delta\hat u(1,t))\\
\Delta \hat u(0,t) &= -\Delta  \hat v(0,t) + q_0(t) - q_{in} - \hat \chi(t)\\
\Delta \hat v(1,t) &= U(t)\\
\dot{\hat{\chi}}(t) &= L(\Delta y(t) - \Delta \hat u(1,t))
\end{align}
\label{eq:linear_observer}
\end{subequations}
where $p_1,p_2$ are observer gains defined as
\begin{subequations}
\begin{align}
p_1(x) &:= -L - \frac{1}{2}\exp\left(\frac{\sigma}{\epsilon}(1-x)\right)\Bigg\{ \sigma I_0\left(\frac{|\sigma|}{\epsilon}\sqrt{1-x^2}\right)\notag \\ & \quad - |\sigma|\sqrt{\frac{1+x}{1-x}}I_1\left(\frac{|\sigma|}{\epsilon}\sqrt{1-x^2}\right) \Bigg\}\notag \\ 
& \quad +\frac{L}{2\epsilon}\int_x^1\exp\left(\frac{\sigma}{\epsilon}(\xi -x)\right)\Bigg\{\sigma I_0\left(\frac{|\sigma|}{\epsilon}\sqrt{\xi^2 -x^2}\right)\notag \\ & \quad - |\sigma|\sqrt{\frac{\xi +x}{\xi - x}}I_1\left(\frac{|\sigma|}{\epsilon}\sqrt{\xi^2 - x^2}\right) \Bigg\}d\xi\\
p_2(x) &:= \frac{1}{2}\exp\left(\frac{\sigma}{\epsilon}(1+x)\right)\Bigg\{ \sigma I_0\left(\frac{|\sigma|}{\epsilon}\sqrt{1-x^2}\right)\notag \\ & \quad - |\sigma|\sqrt{\frac{1-x}{1+x}}I_1\left(\frac{|\sigma|}{\epsilon}\sqrt{1-x^2} \right)\Bigg\}\notag \\ 
& \quad -\frac{L}{2\epsilon}\int_x^1\exp\left(\frac{\sigma}{\epsilon}(\xi +x)\right)\Bigg\{\sigma I_0\left(\frac{|\sigma|}{\epsilon}\sqrt{\xi^2 -x^2}\right)\notag \\ & \quad - |\sigma|\sqrt{\frac{\xi -x}{\xi + x}}I_1\left(\frac{|\sigma|}{\epsilon}\sqrt{\xi^2 - x^2}\right) \Bigg\}d\xi
\end{align}
\end{subequations}
where $I_n$ is the modified Bessel function of the first kind of order $n$ and $L <0$ is a real number, $\epsilon$ is defined in~\eqref{eq:epsilon} and $\sigma$ is given by
\begin{equation}
\sigma = -\frac{F}{2\rho}.
\end{equation}

The output signal $\Delta y = \Delta u(1,\cdot)$ can be constructed from the measured boundary pressure and flow signals $(p_l,q_l)$.

Theorem~8 from~\cite{aamo2015leak} guarantees that the observer~\eqref{eq:linear_observer} produces for the plant~\eqref{eq:linear_plant} state estimates $\Delta \hat u(\cdot,t) \rightarrow \Delta u(\cdot,t)$, $\Delta \hat v(\cdot,t) \rightarrow \Delta v(\cdot,t)$ and parameter estimate $\hat \chi (t) \rightarrow \chi$ exponentially in the $L_2$ sense, and additionally that $ \Delta \hat u(0,t) \rightarrow \Delta u(0,t)$, $\Delta \hat v(0,t) \rightarrow \Delta v(0,t)$.

Using this last fact a leak localisation method is proposed as
\begin{subequations}
\begin{align}
\dot{\hat{\delta}}(t) &= proj_{[0,l]}\{\gamma(p_0(t) - \hat p_0(t))\}\\
\hat p_0(t) &= \frac{\sqrt{\beta \rho}}{A}(\hat u(0,t) - \hat v(0,t)) + p_{out} + \frac{\rho}{A^2}\eta \hat \chi(t)\notag \\ & \quad + \frac{F}{A}\hat \delta(t)\hat \chi(t)
\end{align}
\label{eq:position_estimator}
\end{subequations}
with $\gamma > 0$ an adaptive gain. Theorem~10 from~\cite{aamo2015leak} then guarantees~(in the linear regime) that as long as $F,\chi > 0$, we have that $\hat \delta(t) \rightarrow \delta(0)$. Furthermore, assuming the leak is a point leak at position $z^* \in (0,l)$, Corollary~12 from~\cite{aamo2015leak} tells us that $\hat \delta(t) \rightarrow z^*$.

\subsection{Observer applied to nonlinear plant}
To apply the adaptive observer~\eqref{eq:linear_observer} and position estimator~\eqref{eq:position_estimator} to the plant~\eqref{eq:semilinear_plant} with nonlinear friction terms, we reverse first the change of coordinates~\eqref{eq:final_coordinate_transform} and rewrite the observer in $(u,v)$ coordinates via~\eqref{eq:linearisation_decomposition}, reintroducing the nonlinear source terms. Assuming the distribution describes a point leak, this gives
\begin{subequations}
\begin{align}
\hat u_t(x,t) &= -\epsilon \hat u_x(x,t) - \frac{f}{4DA}|\hat u(x,t) + \hat v(x,t) - \hat \delta'(xl)\hat \chi(t)|\notag \\ & \times (\hat u(x,t) + \hat v(x,t) - \hat \delta'(xl)\hat \chi(t)) \notag \\ &+ \underline p_1(x)(y(t) - \hat u(1,t))\\
\hat v_t(x,t) &= \epsilon \hat v_x(x,t) - \frac{f}{4DA}|\hat u(x,t) + \hat v(x,t) - \hat \delta'(xl)\hat \chi(t)|\notag \\ & \times (\hat u(x,t) + \hat v(x,t) - \hat \delta'(xl)\hat \chi(t))\notag \\ & + \underline p_2(x)(y(t) - \hat u(1,t))\\
\hat u(0,t) &= -\hat v(0,t) + q_0(t) - \hat \chi(t)\\
\hat v(1,t) &= \frac{1}{2}( q_l(t) - \frac{A}{\sqrt{\beta\rho}}(p_l(t) + \rho gh) )\\
\dot{\hat{\chi}}(t) &= L\exp(\frac{lF}{2\sqrt{\beta\rho}})( y(t) - \hat u(1,t))\\
\hat \delta'(xl) &= H(xl - \hat \delta(t)) - 1\\
\dot{\hat{\delta}}(t) &= proj_{[0,l]}\{\gamma (p_0(t) - \hat p_0(t)) \}\\
\hat p_0(t) &=  \frac{\sqrt{\beta\rho}}{A}(\hat u(0,t) - \hat v(0,t)) + \frac{\rho}{A^2}\eta\hat\chi(t)
\end{align}
\label{eq:modified_observer}
\end{subequations}
where $H$ is the Heaviside function, $\underline p_1, \underline p_2$ are given by
\begin{align}
\underline p_1(x) &:= p_1(x)\exp\left(\frac{lF}{2\sqrt{\beta\rho}}(1-x)\right)\\
\underline p_2(x) &:=  p_2(x)\exp\left(\frac{lF}{2\sqrt{\beta\rho}}(1+x)\right)
\end{align}
and the output signal $y = u(1,\cdot)$.
Although the convergence results from~\cite{aamo2015leak} do not necessarily apply to the modified observer~\eqref{eq:modified_observer}, it is of practical interest to study how well the observer detects and quantifies leaks occurring in a pipeline modelled by~\eqref{eq:semilinear_pipe_model}--\eqref{eq:pipeline_boundary}. In the next section simulation results testing the observer at different operating points are presented.

\section{Simulation}\label{sec:simulation}
The semilinear pipeline model~\eqref{eq:semilinear_pipe_model}--\eqref{eq:pipeline_boundary} is implemented in \texttt{MATLAB} with the following parameters
\begin{align*}
l &= 10^3~m  & D &= 0.5~m, & A &= \frac{\pi}{4}(0.5)^2~m^2\\
\beta &= 2.15\times 10^9~Pa, & \rho &= 10^3~\frac{kg}{m^3}, & \nu &= 1.0016~mPa\cdot s
\end{align*}
The inclination angle $\phi$ is set so that the inlet of the pipeline is $10~m$ above the outlet, and the acceleration of gravity is assumed to be $g = 9.8~m^2/s$. Atmospheric pressure of $p_{in} = 10^5~Pa$ is imposed at the inlet. The adaptive gains are set to be
\begin{align*}
L = -1, && \gamma = 0.2.
\end{align*}

\begin{figure}
\centering
\includegraphics[scale=0.36]{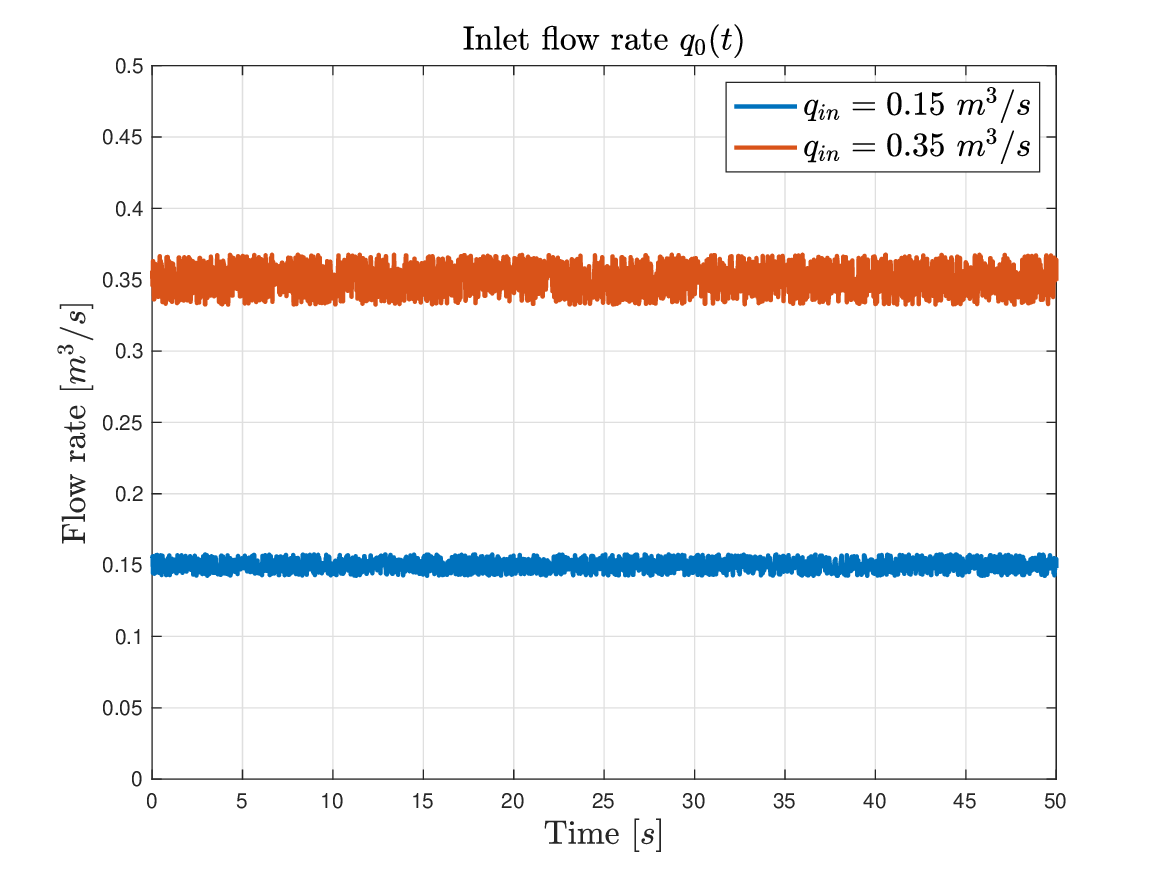}
\caption{Inlet signal $q_0$ for two cases considered.}
\label{fig:inflow}
\end{figure}

The pipeline is simulated for two different mean inlet flows, namely $q_{in} = 0.15,~0.35~m^3/s$, for a duration of $t = 50~s$. Random noise is added to simulate fluctuations in demand around the mean, resulting in the values for the inlet flow signal $q_0$ as plotted in Figure~\ref{fig:inflow}.

%\begin{figure}[ht]
%\begin{subfigure}{.5\textwidth}
  %\centering
  % include first image
  %\includegraphics[width=.8\linewidth]{Simulations/p1}  
  %\caption{Put your sub-caption here}
  %\label{fig:sub-first}
%\end{subfigure}
%\begin{subfigure}{.5\textwidth}
  %\centering
  % include second image
  %\includegraphics[width=.8\linewidth]{Simulations/p2}  
  %\caption{Put your sub-caption here}
  %\label{fig:sub-second}
%\end{subfigure}
%\caption{Put your caption here}
%\label{fig:fig}
%\end{figure}

%\begin{figure}
%\centering
%\includegraphics[scale=0.5]{Simulations/p1}
%\label{fig:inflow}
%\caption{Observer gains $\underline p_1$.}
%\end{figure}
%\begin{figure}
%\centering
%\includegraphics[scale=0.5]{Simulations/p2}
%\label{fig:inflow}
%\caption{Observer gains $\underline p_2$.}
%\end{figure}

\begin{figure}[ht]
\begin{subfigure}{.5\textwidth}
  \centering
  % include first image
  \includegraphics[width=.85\linewidth]{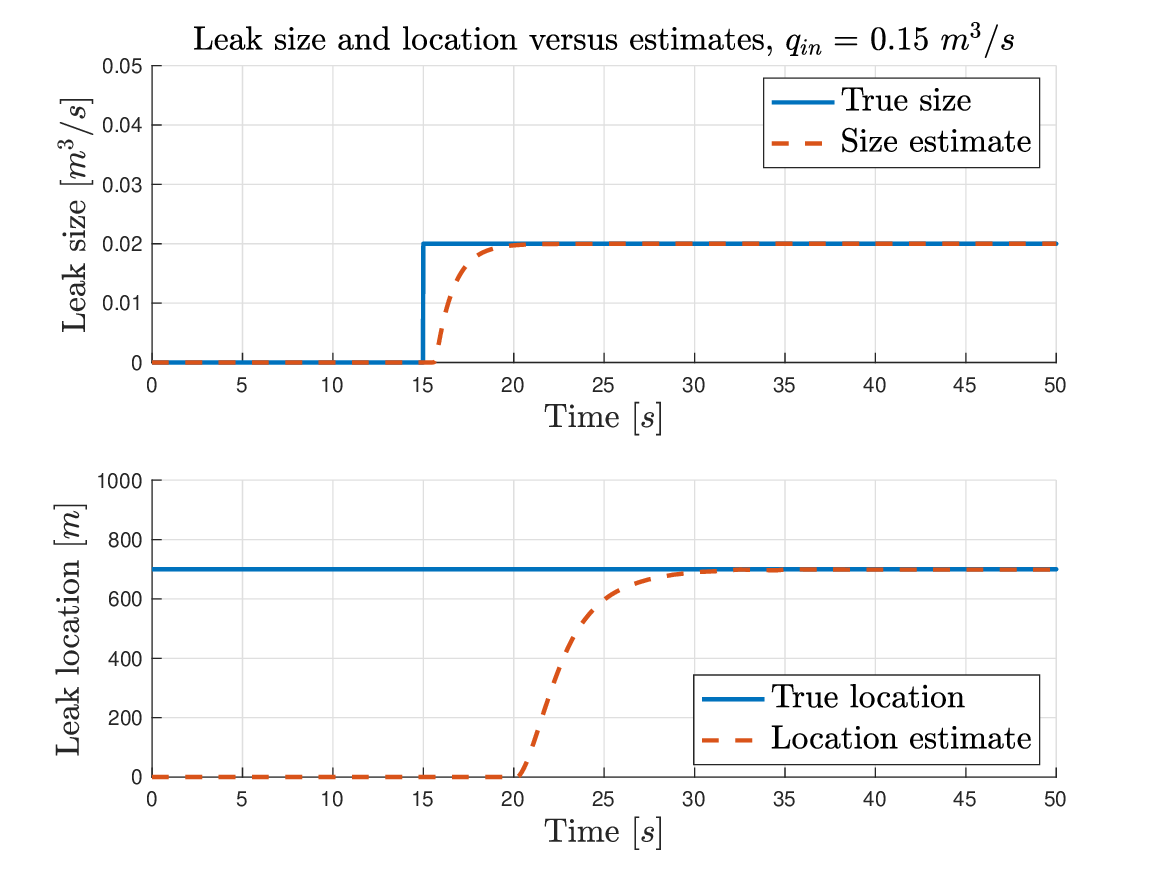}  
  \caption{Size and location estimate for $q_{in} = 0.15~m^3/s$.}
  \label{fig:sub-first}
\end{subfigure}
%\begin{subfigure}{.5\textwidth}
 % \centering
  % include second image
 % \includegraphics[width=\linewidth]{Simulations/q_250}  
  %\caption{Size and location estimate for $q_{in} = 0.25~m^3/s$.}
  %\label{fig:sub-second}
%\end{subfigure}
\begin{subfigure}{.5\textwidth}
  \centering
  % include second image
  \includegraphics[width=.85\linewidth]{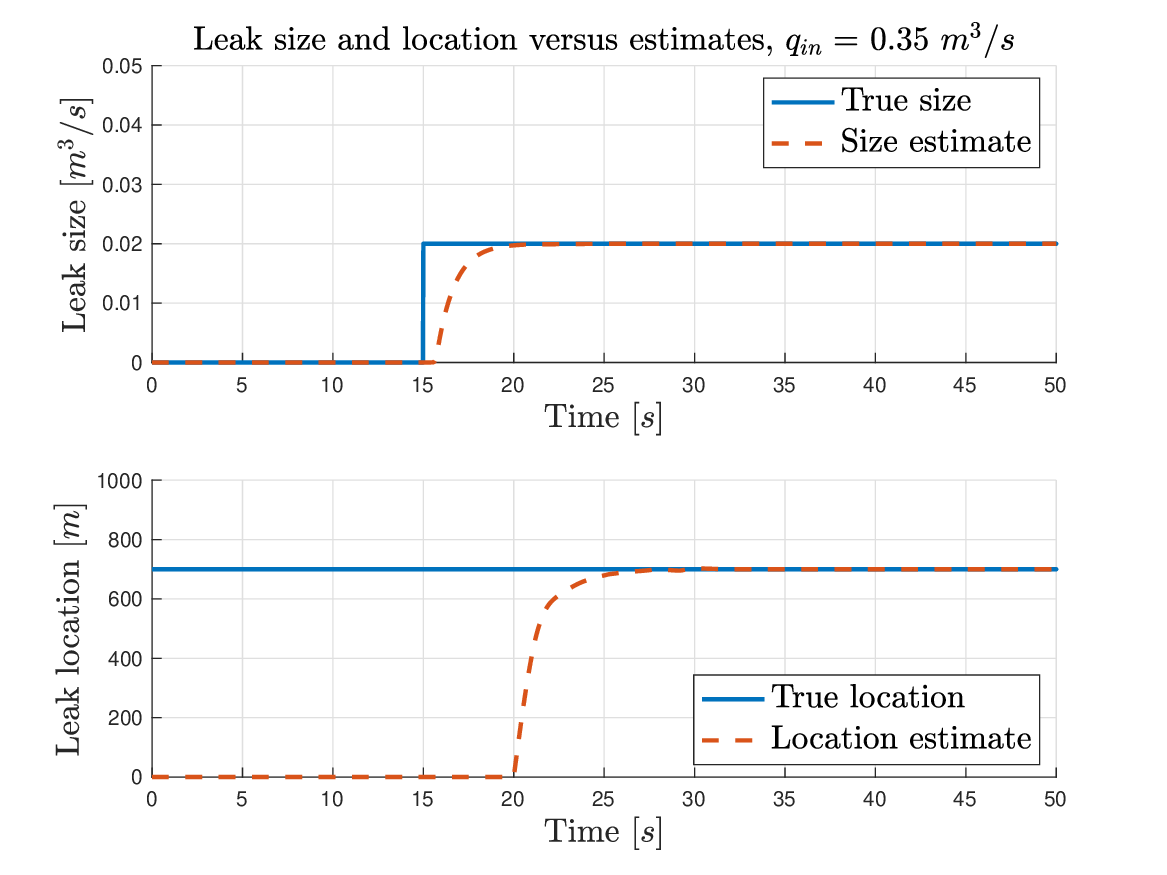}  
  \caption{Size and location estimate for $q_{in} = 0.35~m^3/s$.}
  \label{fig:sub-second}
\end{subfigure}
\caption{Leak detection results.}
\label{fig:leak_detection}
\end{figure}

A point leak of size $\chi = 0.02~m^3/s$ is introduced at $z^* = 700~m$ at time $t^*= 15~s$ for the two cases. The location estimators are turned on $5~s$ after the leaks occur. The plots of the size estimate $\hat \chi(t)$ versus the true size and location estimate $\hat \delta(t)$ versus true location is plotted in Figure~\ref{fig:leak_detection}. Comparing the performance, we see the performance  in size estimation is almost identical between the two cases, but the location estimate converges faster for higher mean flow rates. This reflects that the adaptation rate of location estimate is correlated with the friction in the pipe~(as can be seen from e.g.~\eqref{eq:position_estimator}), which scales with mean flow rate.

\section{Conclusion}
An adaptive observer-based method for detection, size estimation and localisation of leaks in a pipeline based on a linearised, distributed model of the hydraulic transients has been modified to pipelines with nonlinear friction. This increases the application range from only being applicable to pipes with laminar flow to pipes with turbulent flow, which is often the flow regime one is faced with in practice. %It has been shown how the observer gains are calculated for a given operating point, which is typically based on predicted demand levels, and how the observer is to be implemented with nonlinear friction terms. This gives a basis for updating the observer implementation as the demand levels gradually vary on both a daily and seasonal basis.

The simulations in Section~\ref{sec:simulation} show the effectiveness of the proposed leak detection method. %In all three cases the observer successfully estimates the leak size and the leak is localised. 
Intuitively one expects the modified method presented here to inherit convergence properties from the linear design of~\cite{aamo2015leak} due to the inherently stable nature of the plant under consideration. %In particular, the rate at which the observer adapts the leak size estimate to the leak introduced into the system increases for higher friction factor. 
Overall, the proposed approach shows promise for applying distributed observer-based leak detection methods initially designed for linearised pipeline models to settings with more realistic hydraulic models.

One natural direction of further work is to apply the presented observer to a simulation where the operating point is time-varying, and hence the observer gains must be updated throughout. Also, it would be interesting to study whether a stability analysis can be performed on the modified observer presented here, characterising the stability properties more precisely. %In practice there would be model uncertainties which have not been taken into account here, and studying how robust the leak detection method is with respect to these would be valuable further work. Testing the method with laboratory- or field-generated data could complement such a robustness study. Lastly, the modified observer presented here is only for a single pipeline, so modifying the corresponding leak detection methods for networks with branching points~(\cite{anfinsen2022leak}) and loops~(\cite{wilhelmsen2022leak}) in a similar way to account for nonlinear friction is a direction this work will be carried forward.
\bibliography{ifacconf}             % bib file to produce the bibliography
                                                     % with bibtex (preferred)
                                                   
%\begin{thebibliography}{xx}  % you can also add the bibliography by hand

%\bibitem[Able(1956)]{Abl:56}
%B.C. Able.
%\newblock Nucleic acid content of microscope.
%\newblock \emph{Nature}, 135:\penalty0 7--9, 1956.

%\bibitem[Able et~al.(1954)Able, Tagg, and Rush]{AbTaRu:54}
%B.C. Able, R.A. Tagg, and M.~Rush.
%\newblock Enzyme-catalyzed cellular transanimations.
%\newblock In A.F. Round, editor, \emph{Advances in Enzymology}, volume~2, pages
%  125--247. Academic Press, New York, 3rd edition, 1954.

%\bibitem[Keohane(1958)]{Keo:58}
%\newblock \emph{Power and Interdependence: World Politics in Transitions}.
%\newblock Little, Brown \& Co., Boston, 1958.

%\bibitem[Powers(1985)]{Pow:85}
%\newblock Is there a way out?
%\newblock \emph{Harpers}, pages 35--47, June 1985.

%\bibitem[Soukhanov(1992)]{Heritage:92}
%\newblock \emph{{The American Heritage. Dictionary of the American Language}}.
%\newblock Houghton Mifflin Company, 1992.

%\end{thebibliography}

%\appendix
%\section{A summary of Latin grammar}    % Each appendix must have a short title.
%\section{Some Latin vocabulary}              % Sections and subsections are supported  
                                                                         % in the appendices.
\end{document}